\documentclass[10pt,a4paper,superscriptaddress,aps,prd,showkeys,showpacs,nofootinbib,reprint]{revtex4-1}
\usepackage[utf8]{inputenc}
\usepackage{verbatim}
\usepackage{amsmath,amsfonts,amssymb}
\usepackage[breaklinks,colorlinks]{hyperref}
\usepackage{natbib}
\usepackage{tikz}
\usepackage{float}
\usepackage{graphicx}
\usepackage{adjustbox}
\usepackage{tabularx}
\usepackage{amsbsy}
\usepackage{braket}
\hypersetup{%
,urlcolor=blue
,citecolor=blue
,linkcolor=blue
}

\begin{document}

\title{Bending of light in axion backgrounds}

\author{Jamie~I.~McDonald}
\email[]{jamie.mcdonald@tum.de}
\affiliation{%
 Physik-Department, James-Franck-Stra{\ss}e,  Technische Universit{\"a}t M{\"u}nchen,
	85748 Garching, Germany}

\author{Lu{\'i}s B.~Ventura}
\email[]{lbventura@ua.pt}
\affiliation{%
Departamento de F{\'i}sica da Universidade de Aveiro and CIDMA, Campus de Santiago, 3810-183 Aveiro, Portugal}%

\date{\today}

\begin{abstract}
In this work we examine refraction of light by computing full solutions to axion electrodynamics. We also allow for the possibility of an additional plasma component. 
We then specialise to wavelengths which are small compared to background scales to determine if refraction can be described by geometric optics. 
In the absence of plasma, for small incidence angles relative to the optical axis, axion electrodynamics and geometric optics are in good agreement, with refraction occurring at $\mathcal{O}(g_{a \gamma \gamma}^2)$. However, for rays which lie far from the optical axis, the agreement with geometric optics breaks down and the dominant refraction requires a full wave-optical treatment,  occurring at $\mathcal{O}(g_{a \gamma \gamma})$. In the presence of sufficiently large plasma masses, the wave-like nature of light becomes suppressed and geometric optics is in good agreement with the full theory for all rays. Our results therefore suggest the necessity of a more comprehensive study of lensing and ray-tracing in axion backgrounds, including  a full account of the novel $\mathcal{O}(g_{a \gamma \gamma})$ wave-optical contribution to refraction. 
\end{abstract}

\maketitle

\section{Introduction}

Axions \cite{Peccei:1977hh,Weinberg:1977ma,Wilczek:1977pj,Conlon:2006tq, Svrcek:2006yi} remain one of the most promising directions for physics beyond the Standard Model, still offering a viable solution to the dark matter problem \cite{Preskill:1982cy,Abbott:1982af,Dine:1982ah,Marsh:2015xka,Hui:2016ltb,Alonso-Alvarez:2019ssa}. These particles are the subject of many current and upcoming experiments \cite{Asztalos:2009yp,TheMADMAXWorkingGroup:2016hpc,Millar:2016cjp,Majorovits:2016yvk,Anastassopoulos:2017ftl,Irastorza:2011gs,Dobrich:2013mja,Redondo:2010dp,Adler:2008gk} and the response to the results of the XENON1T collaboration \cite{Aprile:2020tmw} shows that axions continue to enjoy widespread interest across the particle physics community. Of particular relevance is the axion coupling to photons $\mathcal{L}_{a\gamma \gamma} = -g_{a \gamma \gamma} a F_{\mu \nu}\tilde{F}^{\mu \nu}/4$ which continues to be a source for many interesting phenomenological proposals \cite{Hertzberg:2020dbk,Levkov:2020txo,Arza:2020eik,Battye:2019aco,Foster:2020pgt,Caputo:2020rnx,Leroy:2019ghm,Carenza:2019vzg,Dessert:2019dos,Battye:2019aco,Chen:2020eer,Arza:2020eik}.

In astrophysical settings, polarisation rotation \cite{Harari:1992ea,Carroll:1989vb,Finelli:2008jv} is one of the most widely studied phenomena concerning photon propagation in axion backgrounds, due to its simplicity and lack of suppression at high frequencies, since the effect is achromatic. This achromaticity has the additional observational advantage of distinguishing the axion-induced rotation from frequency-dependent Faraday rotation \cite{Suresh:2018ayj}. Indeed there continue to be new lessons to learnt from this effect,  \cite{Fedderke:2019ajk,Sigl:2018fba,Sigl:2019pmj,Basu:2020gsy}, including its application to novel axion backgrounds \cite{Agrawal:2019lkr,Chen:2019fsq,Poddar:2020qft,DeRocco:2018jwe,Liu:2019brz,Chigusa:2019rra,Caputo:2020quz}.

There remains, however, a plethora of interesting effects concerning the propagation of photons through axion backgrounds. Perhaps one of the most notable phenomena is \textit{optical} lensing and refraction of light by axion backgrounds due to the axion-photon coupling. Optical lensing, which relies on the local refractive index of the medium, should be distinguished from gravitational lensing arising from long-range gravitational forces (see \cite{Kolb:1995bu,Fairbairn:2017sil,Fairbairn:2017dmf,Kolb:1995bu} and references therein). In particular, the frequency-dependence of optical lensing distinguishes it from its gravitational counterpart which, in the simplest realization, is independent of wavelength due to the strong equivalence principle. 


Furthermore, due to the CP-violating nature of the axion coupling to photons, many of these phenomena are chirality-dependent, \textit{i.e.}, left- and right-circularly polarised photons propagate differently through an axion background. An attempt was made in refs.~\cite{Mohanty:1993zj,Plascencia:2017kca} to compute light refraction in axion backgrounds, where the authors claimed that light was refracted within the eikonal approximation at $\mathcal{O}(g_{a \gamma \gamma})$ \footnote{Since the axion appears derivatively in the photon equations of motion, formally, the appropriate  dimensionless expansion parameter is $\mathcal{O}(k_a g_{a \gamma \gamma} a_0/k_\gamma )$, where $k_{a, \gamma}$ is the typical momentum associated with the axion/photon fields, respectively, $a_0$ is the typical amplitude of the axion field. However, we shall usually write $\mathcal{O}(g_{a \gamma \gamma})$ as a shorthand.}
. However it was later shown in a more systematic treatment  \cite{Blas:2019qqp,McDonald:2019wou} that, within the framework of geometric optics, there is no $\mathcal{O}(g_{a \gamma \gamma})$ refraction of light in a pure axion background. 


Among our previous results \cite{McDonald:2019wou}, we showed that higher order corrections in $g_{a \gamma \gamma}$ to the dispersion relation -- similar to those discussed in \cite{Carroll:1989vb} -- occur at the level of the eikonal equation\footnote{This is the first equation in the hierarchy of geometric optics, from which mass-shell relations in a theory are derived. The next order in gradients, which we did not compute either this or previous work \cite{McDonald:2019wou}, gives a set of transport equations which govern the evolution of field amplitudes along rays.}. These higher order corrections lead to refraction of light at $\mathcal{O}(g_{a \gamma \gamma}^2)$, within the framework of geometric optics. Furthermore, we also showed that, in the presence of a non-trivial background refractive index (provided by a collisionless plasma), polarisation-dependent refraction of light occurred at $\mathcal{O}(g_{a \gamma \gamma})$ within geometric optics.

There is a wide variety of astrophysical axion backgrounds which could give rise to optical lensing described in this paper. These might include axion miniclusters, axion stars and sufficiently dense sub-structures \cite{Eby:2020eas,Eby:2019ntd,Lentz:2019xcr,Braaten:2019knj,Braaten:2018nag,Braaten:2019knj,Braaten:2015eeu,Ellis:2020gtq}, scalar profiles in and around stars \cite{Cardoso:2015zqa,Garbrecht:2018akc, Day:2019bbh, Kaplan:2019ako,Balkin:2020dsr} or superradiant black holes profiles \cite{Wong:2019kru,Arvanitaki:2009fg,Arvanitaki:2014wva,Detweiler:1980uk,Mathur:2020aqv} as suggested in \cite{Plascencia:2017kca,Mohanty:1993zj} or modified theories of gravity involving axions \cite{Odintsov:2020iui,Nojiri:2020pqr}. Another interesting possibility is axion strings \cite{Gorghetto:2018myk}, which have especially large axion field values.


Interestingly, our results have already been applied to optical lensing by axion stars \cite{Prabhu:2020pzm}. In light of the debate concerning the propagation of light signals in axion backgrounds, it is therefore vital to have a reliable account of lensing in axion backgrounds so that any specific phenomenological proposal can be trusted. Any effective description of optics and refraction  must reproduce the predictions of the full theory of axion electrodynamics within its regime of application if it is to be of any use. The purpose of the present paper is to examine to what extent this is achieved by geometric optics as well as quantifying corrections which go beyond the short-wavelength regime.

The remainder of this paper is structured as follows. In sec.~\ref{sec:Maxwell}, we introduce axion electrodynamics, consider the possibility of an additional plasma component and derive a closed set of coupled wave equations in a cold collisionless plasma-axion background. In sec.~\ref{sec:Born}, we find the leading  $\mathcal{O}(g_{a \gamma \gamma})$ analytical solution for the scattered field produced by a monochromatic plane wave incident on a compact axion background at an oblique angle relative to axion gradients. In sec.~\ref{sec:numerics}, we describe a procedure for solving the axion-electrodynamics equations order by order in $g_{a \gamma \gamma}$ for axion backgrounds with harmonic time-dependence. In sec.~\ref{sec:GeometricOptics} we review the geometric optics approximation in an axion-plasma background developed in refs.~\cite{Blas:2019qqp,McDonald:2019wou}. This is used to construct analytic formulae for the refraction angle. The reader not interested in technical details can skip these parts of the paper and move straight to the results of sec.~\ref{sec:results}, where we compare the validity of geometric optics with full axion electrodynamics in the short-wavelength limit. We provide an interpretation of these results in \ref{sec:interpretation} and compare to discussions elsewhere in the literature. Finally sec.~\ref{sec:conclusions} presents our conclusions and proposes some directions for follow-up work.

\section{Axion electrodynamics}\label{sec:Maxwell}
We begin with the equations for axion electrodynamics resulting from the axion-photon coupling
\begin{align}
\nabla \cdot \textbf{E}& = \rho - g_{a \gamma \gamma} \textbf{B} \cdot \nabla  a\,, \label{Gauss}\\
\nabla \times \textbf{B} - \dot{\textbf{E}} &= \textbf{J} +  g_{a \gamma \gamma}\dot{a} \textbf{B} + g_{a \gamma \gamma} \nabla a \times \textbf{E}\,,\label{curlB}\\
\nabla \cdot \textbf{B} &=0 \label{divB}\,, \\
\dot{\textbf{B} } + \nabla \times \textbf{E}& =0, \label{Bianchi2}
\end{align}
where $\textbf{E}$ and $\textbf{B}$ are the electric and magnetic fields, $\textbf{J}$ and $\rho$ are current and charge densities. 


For a cold, collisionless, plasma background with no background electromagnetic fields, the plasma change density $\rho_{\rm p}$ and current $\textbf{J}_{\rm p}$ satisfy
\begin{align}
\dot{\rho}_{\rm p} + \nabla \cdot \textbf{J}_{\rm p} = 0, \label{PlasmaCons} \\
\dot{\textbf{J}}_{\rm p}=\omega_{\rm p}^2 \textbf{E}. \label{PlasmaCurrent} 
\end{align}
The first equation follows from charge conservation in the plasma, whereas the second equation follows from the Lorentz force acting on the plasma with $\omega_{\rm p} = \sqrt{n/m}$, where $\omega_{ \rm p}$ is the plasma frequency, $n$ is the background density of charge carriers and $m$ is the mass of the charge carriers.

Combining the Bianchi identity \eqref{Bianchi2} with either the time-derivative or curl of \eqref{curlB}, we can obtain the standard closed system of \textit{coupled} wave equations for $\mathbf{E}$ and $\mathbf{B}$, respectively, in an axion-plasma background:
\begin{align}
&\square \textbf{E} + \nabla(\nabla \cdot \textbf{E}) + \dot{\textbf{J}}_p + \dot{\textbf{J}}_a =0, \label{EEq}\\
&\square \textbf{B}  -  \nabla \times \textbf{J}_a  - \nabla \times \textbf{J}_{\rm p}=0 \label{BEq},
\end{align}
where $\square = (\partial_t^2 - \nabla^2)$ and the axion and plasma current densities are
\begin{align}
&\textbf{J}_a =  g_{a \gamma \gamma}  \left[\dot{a} \textbf{B} + \nabla a \times \textbf{E} \right], \label{eq:Ja}\\
& 
\dot{\textbf{J}}_{\rm p} = \, \omega_{\rm p}^2  \, \textbf{E}. 
\end{align}

\subsection*{Setup}

\begin{figure}
    \centering
    \includegraphics[scale=0.9]{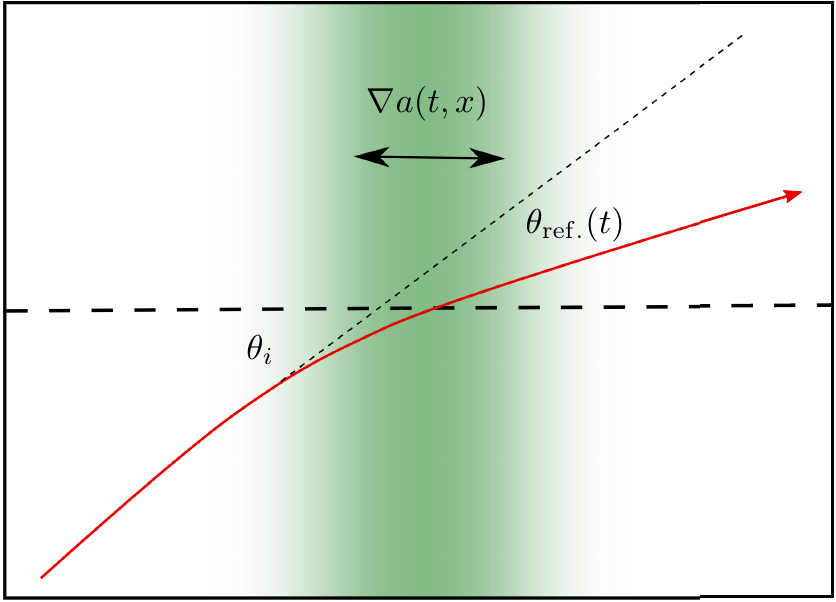}
    \caption{Illustration of the setup used in this paper. A wave-front (red) approaches a compact axion background (green) which goes to zero smoothly at $x \rightarrow \pm \infty$. The incident beam makes an angle of incidence $\theta_i$ relative to the optical axis (black dashed) defined by axion gradients. The ray is then refracted by an angle $\theta_{\rm ref.}$ relative to the initial trajectory. For an oscillating axion background, $\theta_{\rm ref.}$ is time-dependent.}
    \label{fig:RefractionCartoon}
\end{figure}

The aim of this paper is to study refraction of a monochromatic wave passing through a compact axion background.  Therefore, we consider the most minimal case to expose the relevant physics, whilst being simple enough to allow analytic solutions and an easy numerical comparison. In general, solving axion-electrodynamics in an arbitrary geometry is a non-trivial task \cite{Ouellet:2018nfr,Knirck:2019eug,Battye:2019aco}. In addition, if one wishes to carry out numerical simulations at very short wavelengths relevant for the limit of geometric optics, high resolution within the integration volume is required. For generic geometries, this can lead to a rapid increase in total computing time.  However, here we shall consider the simple case where a compact axion profile extending in 1 dimension,
 \begin{equation}\label{eq:aprofile}
 a = a(t,x), \qquad a \rightarrow 0 \quad \text{as} \quad x \rightarrow \pm \infty ,
 \end{equation}
is illuminated by a propagating oblique plane wave
\begin{align}\label{eq:EInc}
    \textbf{E}_{\rm inc}  &=  \boldsymbol{\varepsilon}_0
    e^{- i p^0 t + i \textbf{p}\cdot \textbf{x}  },
\end{align}
where $\boldsymbol{\varepsilon}_0$ is the initial polarisation and $\mathbf{p} = (p^x,0,p^z)$ is the initial wave-vector.
This setup is illustrated in fig.~\ref{fig:RefractionCartoon}. 


The initial wave-vector $\mathbf{p}$ and the normal to the axion surface (see fig.~\ref{fig:RefractionCartoon}) define an incidence angle, $\theta_i$. Thus $\theta_i$ parametrises how close the incident ray is to the direction of gradients of the axion background: if $\theta_i = 0$, the incidence direction is parallel to these gradients. Since the axion background has no $z$-dependence, the momentum $p_z$ perpendicular to axion gradients is conserved: momentum transfer occurs only in the $x$-direction. 

The Poynting flux of an electromagnetic wave
\begin{equation}
\textbf{S} = \textbf{E}\times \textbf{B},
\end{equation}
gives the integral curves along which energy is transported. By computing the angle between the outgoing and incident Poynting flux, one can measure the angle of refraction of the incoming beam. 


\section{Born Approximation}\label{sec:Born}

We now compute the leading $\mathcal{O}(g_{a \gamma \gamma})$ response to the incident beam \eqref{eq:EInc} using the equations of axion electrodynamics. This consists of the decomposition
\begin{equation}
    \textbf{E} = \textbf{E}_{\rm  inc.} + \textbf{E}_{\rm scat.} ,
\end{equation}
where $\textbf{E}_{\rm scat.}$ is the scattered field. A similar decomposition holds for the magnetic field. Since our axion field takes the form $a=a(t,x)$, the $z$-component of the incoming momentum $p_z$ is conserved, and so the solution factorises as
\begin{equation}\label{eq:Eansatz}
 \textbf{E}_{\rm scat.}(t,x,z)  = e^{i p_z z}\bar{\textbf{E}}(t,x),
\end{equation}
for some function $\bar{\textbf{E}}(t,x)$. Thus the problem essentially reduces to determining the 1+1 dimensional dynamics of the reduced field $\bar{\textbf{E}}(t,x)$. By substituting this ansatz into the wave equation \eqref{EEq}, one can see the field $\bar{\textbf{E}}$ is given by
\begin{align}\label{eq:GR}
&\bar{\textbf{E}}(t,x)= - \int dt' \int d x' \textbf{G}_R(t-t',x-x') \cdot   \dot{\bar{\textbf{J}}}_a(t',x'),
\end{align}
where $\textbf{G}_R(t-t',x-x')$ is a retarded Green function defined below, and  $\bar{\textbf{J}}_a(t,x)$ follows from the same decomposition as eq.~\eqref{eq:Eansatz}.
Explicitly, $\textbf{G}_R(t-t',x-x')$ is a reduced 1+1 dimensional retarded Green function satisfying
\begin{equation}
\boldsymbol{G}_R(t,x) = \int \frac{d k^0}{2 \pi}\int \frac{d k_x}{2 \pi} \tilde{\textbf{G}}_R(k^0,k_x)e^{- i k^0 t + i k_x x},
\end{equation}
where $\textbf{k} =(k_x,0,p_z)$ and $\tilde{\textbf{G}}_R(k^0,\textbf{k})$ obeys
\begin{equation}
\left[ - (k^0)^2 + \left|\textbf{k}\right|^2 - \textbf{k} \otimes \textbf{k} + \omega_{\rm p}^2 \right] \tilde{\textbf{G}}_R(k^0,k_x) = 1.
\end{equation}
By inverting this equation, one arrives at the following expression for the Green function
\begin{align}
&\boldsymbol{G}_R(t,x) \nonumber \\
&= - \int \frac{d k^0}{2 \pi}\int \frac{d k_x}{2 \pi}\frac{\left[(k^0)^2- \omega_{\rm p}^2 - \textbf{k} \otimes \textbf{k} \right]e^{- i k^0 t + i k_xx}}{ \left[ (k^0)^2 - \omega_{\rm p}^2 \right]\left[(k^0)^2- \left|\textbf{k}\right|^2 - \omega_{\rm p}^2 \right]} .
\end{align}
To proceed, we must choose a contour for the $k^0$ integral. This is done by ensuring outgoing waves at $\left| x \right| \rightarrow \infty$. The contour then encloses only those poles which have right-moving waves at $x \rightarrow \infty$ and left-moving waves at $x \rightarrow - \infty$. Note we do not pick the residues at $k^0 = \omega_{\rm p}$ as these correspond to zero mode solutions. In addition, we impose causality such that the Green function is non-vanishing only inside the light cone: $ |x| \leq t$. This leads to
\begin{align}\label{eq:GRExplicit}
\boldsymbol{G}_R(t,x)=& - i \theta\left( t - |x|\right)
\int_{-\infty}^{\infty}  \frac{d k_x}{4 \pi \sigma(k_x)\omega(\textbf{k})}\left[\mathbf{1} - \frac{\textbf{k} \otimes \textbf{k}}{|\textbf{k}|^2} \right] \nonumber \\
&\cdot  e^{- i \sigma(k_x) \omega t +  i k_x \left|x\right|},
\end{align}
where $\omega(k) = \sqrt{k_x^2+k_z^2 + \omega_{\rm p}^2 }$ is the on-shell energy and $\sigma$ is the sign function, which ensures the phase velocity at the boundaries is always outgoing. We can then substitute the form of the Green function \eqref{eq:GRExplicit} into eq.~\eqref{eq:GR}, which gives
\begin{align}
&\bar{\textbf{E}}(t,x) =  i  \int _{-\infty}^t dt' \int^{x+t}_{x-t} dx'  \int \frac{d k_x}{4 \pi \sigma (k_x) \omega(\textbf{k})} \nonumber \\
&\cdot \left[\mathbf{1} - \frac{\textbf{k} \otimes \textbf{k}}{|\textbf{k}|^2} \right]   e^{- i \sigma(k_x) \omega (t-t')+ i  k_x \left|x - x'\right|}  \dot{\bar{\textbf{J}}}_a(t',x').
\end{align}
For the purposes of understanding refraction, we are only interested in steady-state solutions far from the axion source, and it is enough to study the scattered wave for large $t$ and $x$, obeying $t \geq |x|$. This gives the transmitted part of the scattered field defined by
\begin{align}
\bar{\textbf{E}}^T(t,x)=\lim_{t,x \rightarrow \infty} \bar{\textbf{E}}(t,x),
\end{align}
so that
\begin{align}\label{eq:ETRealSpace}
&\bar{\textbf{E}}^T(t,x) =  i \int _{-\infty}^\infty dt' \int^{\infty}_{-\infty} dx'  \int \frac{d k_x}{4 \pi \sigma (k_x) \omega(\textbf{k})} \nonumber \\
&\cdot \left[\mathbf{1} - \frac{\textbf{k} \otimes \textbf{k}}{|\textbf{k}|^2} \right]   e^{- i \sigma(k_x) \omega (t-t')+ i k_x \left|x - x'\right|}  \dot{\bar{\textbf{J}}}_a(t',x').
\end{align}
Since we are only computing the first order Born approximation, the axion current appearing in the above equation is evaluated with the incident electromagnetic fields given by \eqref{eq:EInc}. Given the infinite integration ranges, we can now re-write \eqref{eq:ETRealSpace} in terms of Fourier transforms as
\begin{align}
  \bar{\textbf{E}}^T(t,x) = &    \int_{-\infty}^{\infty} \frac{d k_x}{4 \pi}\Bigg[1 - \frac{\textbf{k} \otimes \textbf{k}}{\left| \textbf{k}\right|^2} \Bigg] e^{- i \epsilon(k_x) \omega t +  i k_x x} \nonumber \\
    &\cdot  \tilde{\textbf{J}}_a( \sigma(k_x)\omega,k_x), \label{IntSolEScat}
\end{align}
where $\tilde{\textbf{J}}_a( k^0 , k_x)$ is the Fourier transform of the axion current $\bar{\textbf{J}}_a(t,x)$, evaluated at the monochromatic incident field. Inserting eq.~\eqref{eq:EInc} into eq.~\eqref{eq:Ja}, we compute the Fourier transform
\begin{align}
&\tilde{\textbf{J}}_a(\omega,k_x) = \nonumber \\
&i g_{a \gamma \gamma } ~ \omega ~ \tilde{a}(\omega - p^0, k_x-p_x)\left[\left( \frac{\textbf{p}}{p^0} - \frac{\textbf{k}}{\omega} \right) \times \boldsymbol{\varepsilon}_0 \right]. \label{MomSAxCurr}
\end{align}
where $\varepsilon_0$ is the incident polarisation and the Bianchi identity \eqref{Bianchi2} was used to write the incident magnetic field as $\textbf{B}_{\rm inc.} = \left(\textbf{p}/p^0\right) \times \textbf{E}_{\rm inc.}$. The axion field therefore provides the momentum and energy transfer between incoming and outgoing states. Combining \eqref{IntSolEScat} and \eqref{MomSAxCurr}, the transmitted scattered field can be written as
\begin{align}\label{eq:ETrans}
    &\bar{\textbf{E}}^T(t,x) =  i g_{a \gamma \gamma}   \int_{-\infty}^{\infty} \frac{d k_x}{4 \pi}\tilde{a}
    (\sigma(k_x)\omega - p^0, k_x-p_x) \nonumber \\
    & \cdot \sigma(k_x)\omega(\textbf{k}) \Bigg[1 - \frac{\textbf{k} \otimes \textbf{k}}{\left|\textbf{k}\right|^2} \Bigg] \cdot \Bigg[\left( \frac{\textbf{p}}{p^0} - \frac{\textbf{k}}{\omega(\textbf{k})} \right) \times \boldsymbol{\varepsilon}_0 \Bigg]\nonumber \\
    &\cdot e^{- i \sigma(k_x) \omega t +  ik_x x}.
\end{align}
We then specialise to fields which have a simple harmonic time-dependence
\begin{equation}\label{eq:NRAxion}
a(t,x) = e^{- i m_a t} \psi(x)+ e^{ i m_a t} \psi^*(x),
\end{equation}
for some spatial profile $\psi(x)$. In this case, the Fourier transform of the axion field is
\begin{equation}\label{eq:aFT}
\tilde{a}(q^0, q_x) = 2\pi\left[ \delta(q^0-m_a) \tilde{\psi}(q_x) +\delta(q^0+ m_a ) \tilde{\psi}^*(q_x) \right],
\end{equation}
where $\tilde{\psi}(q_x) = \int d x \psi(x) e^{i q_x \cdot x}$ is the Fourier transform of the axion spatial profile $\psi(x)$.

Next we must use the delta function to impose momentum conservation, thereby selecting the kinematically allowed processes. One might expect some interesting threshold behaviour, and perhaps even resonances for the regime $m_a \geq p^0$. 
However, in the present paper, we are mainly interested in the short wavelength limit, since in particular, as we want to compare the full results of axion electrodynamics to geometric optics results of our previous work \cite{McDonald:2019wou}. We therefore leave these points of interest for future studies, and for the remainder of the paper work only in the regime $m_a < p^0$. In this frequency range, there are two outgoing modes with 4-momenta
\begin{align}\label{eq:PLR}
p^\pm = \left((p^0 \pm m_a), \sqrt{(p^0 \pm m_a)^2 - \omega_{\rm p}^2-p_z^2} , 0 , p_z   \right). 
\end{align}
The electric and magnetic fields are then given by
\begin{equation}\label{eq:EandBBOrn}
\textbf{E}^{T}_{\rm scat.} = -\dot{\textbf{A}}^T_{\rm scat.}, \qquad \textbf{B}^{T}_{\rm scat.} = \nabla \times \textbf{A}^T_{\rm scat.},
\end{equation}
From eq.~\eqref{eq:ETrans}, one can infer
\begin{equation}
\textbf{A}^T_{\rm scat.} =  \textbf{A}(p^+,\psi) + \textbf{A}(p^- , \psi^*),
\end{equation}
and $\textbf{A}$ is a function of the incoming and outgoing 4-momenta $p_\mu$ and $p'_\mu$ given by
\begin{widetext}
\begin{align}\label{eq:ABorn}
 \textbf{A} (p', \psi) = \frac{  g_{a \gamma \gamma}}{2} \Bigg[1- \frac{\textbf{p}' \otimes \textbf{p}' }{\left| \textbf{p}'\right|^2}\Bigg] \cdot \Bigg[ \left(\frac{\textbf{p}' }{p'^{\,0}}- \frac{\textbf{p}}{p^0} \right) \times  \boldsymbol{\varepsilon}_0 \Bigg] \frac{p'^{\,0} }{p_x'  }  \tilde{\psi}\left(   p'_x  -  p_x  \right) e^{- i p'^{\,0} t  + i \textbf{p}'\cdot \textbf{x}},
\end{align}
\end{widetext}
where $p' = p^{\pm}$, eq.~\eqref{eq:PLR}. This structure can be seen to follow from eqs.~\eqref{eq:ETrans} and ~\eqref{eq:aFT} after taking zeros of the delta function with respect to $k_x$. 
Note that these expressions have re-incorporated the overall conserved-momentum factor $e^{i p_z z}$. Thus, $p^\pm$ are just on-shell momenta which have picked up a frequency shift $\pm m_a$.

\section{Numerical Solutions}\label{sec:numerics}


The result \eqref{eq:ABorn} is actually only the first contribution in a power series expansion in $g_{a \gamma \gamma}$, which can be derived iteratively to construct solutions up to any order in $g_{a \gamma \gamma}$. Furthermore, by virtue of the time-harmonic nature of the axion field and incident waves, and the linearity of the equations, the accounting procedure for different orders actually becomes straightforward. 

We shall construct solutions as a Born series in $g_{a \gamma \gamma}$
 \begin{equation}
     \textbf{E} = \sum_{n=0}^\infty \textbf{E}^{(n)}, 
 \end{equation}
where $\textbf{E}^{(n)}$ is the $\mathcal{O}(g_{a \gamma \gamma}^n)$ contribution and $\textbf{E}^{(0)}$ is the incident wave. Thus, from eqs.~\eqref{EEq} and \eqref{eq:Ja}, we have
\begin{equation}
\begin{aligned}
\square \textbf{E}^{(n)} + \nabla(\nabla \cdot \textbf{E}^{(n)}) + \omega_{\rm p}^2 \textbf{E}^{(n)}  \\
= - g_{a \gamma \gamma}  \partial_t \left(\dot{a} \textbf{B}^{(n-1)} + \nabla a \times \textbf{E}^{(n-1)}\right).
\end{aligned}
\end{equation}
where $\textbf{B}^{(n)}$ is the corresponding $n$-th order magnetic field. This is similar to the procedure outlined in \cite{Knirck:2019eug,Kim:2018sci,Battye:2019aco}. If we now specialise to an axion field with harmonic time-dependence as 
\begin{equation}\label{eq:NRAxionField}
a(t,\textbf{x}) = e^{- i m_a t} \psi(\textbf{x})+ e^{ i m_a t} \psi^*(\textbf{x}),
\end{equation}
one can solve the tower of equations in frequency space order by order in $g_{a \gamma \gamma}$.



\begin{figure}[t!]
\includegraphics[scale=1.2]{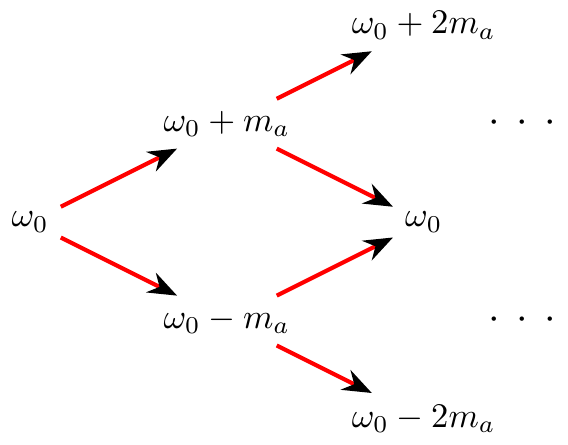}
\caption{Illustration of the iterative procedure used to derive numerical solutions as a Born series in $g_{a \gamma \gamma}$ for an incident wave and axion field with frequencies $\omega_0$ and $m_a$, respectively. Each node gives the frequency of the corresponding mode with the order in $g_{a \gamma \gamma}$ increasing from left to right.}
\label{fig:IterationProcedure}
\end{figure}

If the incident wave has frequency $\omega_0$, then the first order perturbation (the Born approximation) will have a solution with two components, of frequencies $\omega_0 \pm m_a$, as can be seen from eq.~\eqref{eq:PLR}-\eqref{eq:ABorn}. Each of these will source the $\mathcal{O}(g_{a \gamma \gamma}^2)$ solution. Using the linearity of the equations, one can solve for the response field of each of these two modes separately. Thus, by virtue of linearity, solutions can be computed mode-by-mode, order-by-order. This produces at $\mathcal{O}(g_{a \gamma \gamma}^2)$ three modes with frequencies $\omega_0 \pm 2m_a$ and $\omega_0$. This procedure can be iterated to produce solutions to any order in $g_{a \gamma \gamma}$. It is extremely efficient if one is interested only in steady-state solutions as it does away with the need for computationally intensive solutions in the time-domain, allowing easy numerical parameter scanning. The procedure is illustrated in fig.~\ref{fig:IterationProcedure}.

To summarize, the $n$-order solution is given by a sum of $n+1$ harmonics, labelled by the index $m$
\begin{equation}\label{eq:Enplus1}
\textbf{E}^{(n)} (t,\textbf{x})  = \sum_{m}  \bar{\textbf{E}}^{(n)}_m(\textbf{x})\, e^{- i\, \omega^{n}_m t},
\end{equation}
where the $n+1$ order frequencies $\omega^{n+1}_{m}$ are related to the $n$ order frequencies by adding and subtracting appropriate integer multiples of $m_a$ as pictured in fig.~\ref{fig:IterationProcedure}.
Following this approach, by combining  Eqs.~\eqref{eq:NRAxionField} and \eqref{eq:Enplus1}, we need only solve a set of \textit{spatial} equations: 
\begin{align}
& \left[ - \nabla^2 + \omega_{\rm p}^2- (\omega^{n}_{m})^2\right] \textbf{E}^{(n)}_{m }  + \nabla(\nabla \cdot \textbf{E}^{(n)}_{m})   \nonumber \\
&=  \sum_{\pm} g_{a \gamma \gamma} \omega^{n}_{m}\left( \pm i  m_a \psi^\pm \textbf{B}^{(n-1)}_{m\mp 1} + i\nabla \psi^\pm \times \textbf{E}^{(n-1)}_{m \mp 1}\right), \label{IniObOEq} 
\end{align}
where $\psi^+ = \psi$ and $\psi^- = \psi^*$ is a convenient short-hand. Here $\omega^n_{m}$ is the frequency of the order $n$ mode given by adding/subtracting $m_a$ to the appropriate $(n-1)$ order frequencies. Note also that there is some degeneracy, in that one frequency at $\mathcal{O}(g_{a \gamma \gamma}^{n})$ can be sourced by up to two-different modes at $\mathcal{O}(g_{a \gamma \gamma}^{n-1})$, as is apparent from fig.~\ref{fig:IterationProcedure}.


This equation can be further simplified by using the conservation of plasma current $\dot{\rho}_{\rm p} + \nabla \cdot \textbf{J}_{\rm p} =0$ and the relation $\dot{\textbf{J}}_{\rm p} = \omega_{\rm p}^2 \textbf{E}$  in combination with Gauss' law $\nabla \cdot \textbf{E} = \rho_{\rm p} - g_{a \gamma \gamma} \nabla a \cdot \textbf{B}$, leading to the following divergence of the electric field
\begin{equation}
    \nabla \cdot \textbf{E}^{(n)}_{m} = - g_{a \gamma \gamma} \left( 1 - \frac{\omega_{\rm p}^2}{(\omega_m^{n})^2}\right)^{-1} \sum_{\pm} \nabla \psi^\pm \cdot \textbf{B}^{(n-1)}_{m \mp 1}.  
\end{equation}
Replacing this in \eqref{IniObOEq}, one obtains the final form
\begin{widetext}
\begin{align}\label{eq:nOrder}
    &\left[- \nabla^2 + \omega_{\rm p}^2 - (\omega^{n+1}_{m})^2 \ \right] \textbf{E}^{(n)}_{m}\nonumber \\
    &=  \sum_{\pm} g_{a \gamma \gamma} \left[ \left( 1 - \frac{\omega_{\rm p}^2}{(\omega_m^{n})^2}\right)^{-1}  \nabla( \nabla \psi \cdot \textbf{B}^{(n-1)}_{m\mp1}) + \omega^n_{m}\left( \pm i m_a \psi^\pm \textbf{B}^{(n-1)}_{m\mp1} + i\nabla \psi^\pm \times \textbf{E}^{(n-1)}_{m\mp1}\right)  \right]. 
\end{align}
\end{widetext}
Since we have not made any assumptions about the spatial nature of the axion background, this equation can be used in an arbitrary geometry, as long as the axion field and incident beam have harmonic time-dependence. Note the procedure is even simpler when the axion background is stationary. 

\begin{figure}
    \centering
    \includegraphics[scale=0.9]{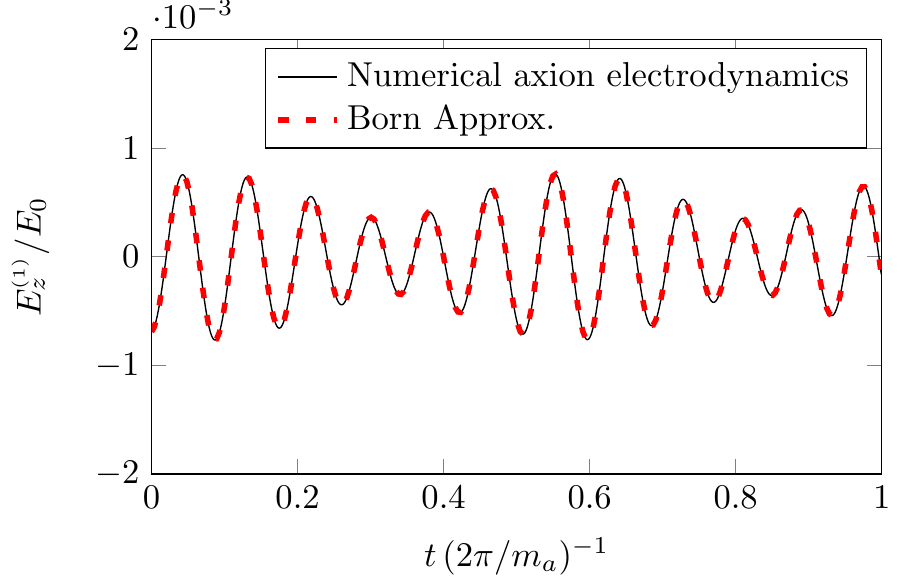}
      \caption{Comparison of the Born approximation \eqref{eq:EandBBOrn}-\eqref{eq:ABorn} and numerical axion-electrodynamics result at $\mathcal{O}(g_{a \gamma \gamma})$. We took an axion field of the form \eqref{eq:NRAxion} with Gaussian spatial profile $\psi(x) = e^{-x^2/L_a^2}$. The axion values in the plot are $m_a = 0.1 \omega_0$, and $L_a \omega_0 = 8$, $g_{ a \gamma \gamma} a_0 = 10^{-2}$. The incidence angle was $\theta_i = \pi/3$ and the plasma frequency was $\omega_{ \rm p} = 0.3 \omega_0$. We chose an incident polarisation $\boldsymbol{\varepsilon}_{0}=(-\sin \theta_i,\, i,\, \cos \theta_i)E_0$.}
    \label{fig:NumericsVsBorn}
\end{figure}

Returning to the oblique scattering problem of sec.~\ref{sec:Born}, the conserved $p_z$ momentum means one can write all solutions as
\begin{equation}
\textbf{E}^{(n)}(x,z) = e^{i p_z z} \bar{\textbf{E}}^{(n)}(x)
,
\end{equation}
and eq.~\eqref{eq:nOrder} becomes a one-dimensional equation in $x$, with boundary condition
\begin{align}
&\frac{d \,\bar{\textbf{E}}^{(n)}_m}{dx}  = \pm
\left[
\left(\omega^{n}_m\right)^2 - p_z^2 
- \omega_{\rm p}^2 \right]^{1/2} \bar{\textbf{E}}^{(n)}_m, \, x\rightarrow \pm \infty.
\end{align}
The corresponding numerical solutions computed from \eqref{eq:nOrder} are shown to agree with the analytic Born approximation \eqref{eq:ABorn} in fig.~\ref{fig:NumericsVsBorn}. We then use these numerical solutions to construct the scattered field of full-axion electrodynamics up to $\mathcal{O}(g_{a \gamma \gamma}^2)$ with results given in sec.~\ref{sec:results}.

\section{Geometric Optics}\label{sec:GeometricOptics}
Here we briefly review the treatment of geometric optics for photons propagating in axion backgrounds, first outlined in \cite{Blas:2019qqp}, and extended in ref.~\cite{McDonald:2019wou}. We begin with the wave equation for the electric field
\begin{align}\label{eq:EWave2}
\square \textbf{E} + \nabla(\nabla \cdot \textbf{E}) + \omega_{\rm p}^2 \textbf{E} + g_{a \gamma \gamma} \partial_t \left[\dot{a}  \textbf{B} + \nabla a \times \textbf{E} \right] =0.
\end{align}
Next we make a geometrical optics approximation, which involves working in the limit in which photon wavelengths are much shorter than background scales. This consists of dropping those terms in eq.~\eqref{eq:EWave2} containing two derivatives of the axion field. This is the standard approximation made in refs \cite{Harari:1992ea,Blas:2019qqp,Carroll:1989vb} and indeed by ourselves in \cite{McDonald:2019wou}. With this approximation, eq.~\eqref{eq:EWave2} reads 
\begin{align}\label{eq:EWaveWKB}
\square \textbf{E} + \nabla(\nabla \cdot \textbf{E}) + \omega_{\rm p}^2 \textbf{E} - g_{a \gamma \gamma} \dot{a} \nabla \times \textbf{E} + g_{a \gamma \gamma}  \nabla a \times \dot{\textbf{E}} \simeq 0.
\end{align}
where we used $\dot{\textbf{B}} = - \nabla \times \textbf{E}$ to eliminate $\textbf{B}$ from the equation.  We then use the limit where the wavelength of the photon is much smaller than the other physical scales of the problem, allowing us to define solutions of the form
\begin{equation}\label{Sols}
\textbf{E} = \textbf{E}_0 \, e^{i S}, \qquad \textbf{B}= \textbf{B}_0 \, e^{i S},
\end{equation}
where frequency and momentum are identified along rays as
\begin{equation}
\omega = -  \dot{S}, \qquad  \textbf{k} = \nabla S.
\end{equation}
From eq.~\eqref{eq:EWaveWKB} we then obtain the following expression
\begin{align}
&\left(\omega^2 - \left|\textbf{k}\right|^2 - \omega_{\rm p}^2 \right)\textbf{E}_0 + \textbf{k}\left( \textbf{k}\cdot \textbf{E}_0\right) \nonumber \\
&+ i g_{a \gamma \gamma}  \dot{a} \, \textbf{k} \times \textbf{E}_0 + i \omega g_{a \gamma \gamma} \nabla \times \textbf{E}_0 =0.
\end{align}
where we have dropped derivatives on $\textbf{E}_0$, which corresponds to the evolution of the field amplitude. Formally, this evolution is captured by a set of transport equations which we do not compute here. 

The operator acting on $\textbf{E}_0$ must have vanishing eigenvalues to satisfy this homogeneous equation. From this we can read off the dispersion relation for modes derived in our previous work \cite{McDonald:2019wou}. This condition requires the vanishing of the quantity
\begin{widetext}
\begin{equation}
D^{\pm}= k^2 - \omega_{\rm p}^2 \pm
\frac{1}{[\omega^2 - \omega_{\rm p}^2]^{1/2}}\Bigg[
\omega^2 g_{a \gamma \gamma}^2 \left(   (k\cdot \partial a)^2 -k^2 (\partial a)^2\right) +\omega_{\rm p}^2 g_{a \gamma \gamma}^2
\left(
\dot{a}^2 k^2 - 2 \dot{a} \omega (k \cdot \partial a) +(\partial a)^2\omega^2
\right)
\Bigg]^{1/2}, \label{CompleteDispRel}
\end{equation}
\end{widetext}
so that the dispersion relation is given by $D^\pm =0$, where the $\pm$ depends on whether light is right or left-circular polarized. From this we can trace rays via a system of Hamiltonian optics equations \cite{WeinbergWKB,Blas:2019qqp}
\begin{align}
\frac{d \textbf{x}}{dt} &= - \frac{\partial D^\pm/\partial \textbf{k}}{\partial D^\pm/\partial \omega}  =\frac{\partial \omega}{\partial \textbf{k}}, \label{dxdt}\\
\frac{d \textbf{k}}{dt}& =  \frac{\partial D^\pm/\partial \textbf{x}}{\partial D^\pm/\partial \omega}  =-\frac{\partial \omega}{\partial \textbf{x}},\label{dkdt}\\
\frac{d \omega}{dt} &= - \frac{\partial D^\pm/\partial t}{\partial D^\pm/\partial \omega} = \frac{\partial \omega}{\partial t}.\label{domegadt}
\end{align}
We can also solve the dispersion relation \eqref{CompleteDispRel} perturbatively in $g_{a \gamma \gamma}$ to arrive at
\begin{align}\label{eq:OmegaSquared}
&\omega^\pm(\textbf{k})  =  \left|\textbf{k}\right| \pm \frac{g_{a\gamma \gamma}}{2} \left[ \hat{\textbf{k}} \cdot \nabla a  + \dot{a} \right] \mp g_{a \gamma \gamma}\dot{a}\frac{\omega_{\rm p}^2}{4\left|\textbf{k}\right|^2} \nonumber \\
&- \frac{g_{a \gamma \gamma}^2}{8 \left|\textbf{k}\right|} \left[\dot{a}^2 + (\hat{\textbf{k}}\cdot \nabla a)^2 - 2|\nabla a|^2 \right]  + \mathcal{O}(g_{a \gamma \gamma}^3) . 
\end{align}
Using the expansion \eqref{eq:OmegaSquared}, we infer the final deflection angle from the orientation of the outgoing momentum $\textbf{k}$ given by integrating \eqref{dkdt} as carried out in \cite{McDonald:2019wou}. Note that this works because, away from the axion profile, group velocity and momentum are parallel. Being careful to expand about the incident momentum $\textbf{p}$, we then find the following formulae for the deflection angle for a photon passing through a compact axion background. In the absence of plasma, for a $2+1$ dimensional background, we have \cite{McDonald:2019wou}
\begin{equation}\label{eq:theta}
\sin \theta_{\rm ref.} =-\frac{g_{a \gamma \gamma}^2}{8 |\textbf{p}|^2} \int_{-\infty}^\infty\, dt\, \nabla_\perp \, \left[(\partial a)^2\right], 
\end{equation}
where $\nabla_\perp$ indicates a derivative taken normal to the direction of the unperturbed ray. By contrast, when plasma is present, geometric optics predicts polarisation-dependent refraction at $\mathcal{O}(g_{a \gamma \gamma})$
\begin{equation}\label{eq:thetaPL}
\sin \,\theta_{\rm ref.} = \pm \frac{g_{a \gamma \gamma}}{2|\textbf{p}|} \int_{-\infty}^{\infty} dt\left[   n_0 \nabla_\perp \dot{a}+ (\hat{\textbf{p}} \cdot \nabla) \nabla_\perp a  \right], 
\end{equation}
where $\hat{\textbf{p}}$ is the unit vector associated to $\textbf{p}$ and
\begin{equation}
    n_0 = \frac{\left|\textbf{p}\right|}{\sqrt{\omega_{\rm p}^2 + \left|\textbf{p}\right|^2}},
\end{equation}
is the refractive index, evaluated for the unperturbed reference ray. In the latter case, the $\pm$ sign corresponds to the differential refraction angle experienced by right/left-circular polarised mode. In both scenarios, $a = a(t,\textbf{x}_0(t))$ is evaluated along an unperturbed reference ray with coordinates $\textbf{x}_0(t)$ and momentum $\textbf{k}_0$. The quantity $\nabla_\perp$ indicates the spatial derivative perpendicular to the direction of the unperturbed reference ray. 



For backgrounds of the form \eqref{eq:NRAxion} and the oblique ray setup illustrated in fig.~\ref{fig:RefractionCartoon}, the refraction angles with and without plasma follow from \eqref{eq:thetaPL}
\begin{align} \label{eq:gRefraction}
    &\sin \theta_{\rm ref.} = \nonumber \\
    &\frac{g_{a \gamma \gamma} m_a^2 \sin \theta_i}{2 \left| \textbf{p} \right| \cos^2 \theta_i} \left[\frac{n_0^2 - 1}{n_0^3}\right] \left|\tilde{\psi}\left(\frac{m_a}{v^{x}_g}\right) \right| \sin(m_a t + \alpha_0),
\end{align}
where $\tilde{\psi}(k) \equiv |\tilde{\psi}(k)| \exp(-i \alpha_0)$ is the Fourier transform of $\psi(x)$ and from \eqref{eq:theta}
\begin{align}\label{eq:gSquaredRefraction}
      \sin \theta_{\rm ref.} &= \frac{g_{a \gamma \gamma}^2 m_a \sin \theta_i}{4 \left| \textbf{p} \right|^2 \cos^2 \theta_i} \Bigg\{ m_a^2 \left|F\left(\frac{2 m_a}{v_g^x}\right)\right|\sin(2 m_a t + \alpha_1)\nonumber \\
      & + \left|G\left(\frac{2 m_a}{v_g^x}\right)\right|\sin(2 m_a t + \alpha_2) \Bigg\},
\end{align}
respectively, where 
\begin{align}
    F(k) &= \int dk' \tilde{\psi}(k') \tilde{\psi}(k-k'), \\
    G(k) &= - \int dk' k' (k-k') \tilde{\psi}(k') \tilde{\psi}(k-k')
\end{align}
are convolutions arising from products of $\psi$ and its derivatives,
and $\alpha_1$ and $\alpha_2$ are phases extracted from these transforms. Here $v^x_g$ is the incident group velocity in the $x$-direction given by $v_g^x = p_x/p^0$.  If the characteristic spatial scales of the axion are $\sim L_a$, it is easy to see after some naive power counting that the Fourier transform scales as $\tilde{\psi} \propto L_a$ and, since the characteristic momentum scales as $k \propto m_a$, $F(k) \sim m_a L_a^2 = m_a^{-1}(L_a m_a)^2$ and $G(k) \sim m_a^3 L_a^2 = m_a (L_a m_a)^2$. Hence, when the product $L_a m_a $ is fixed, we see that \eqref{eq:gSquaredRefraction} scales as $m_a^2$ for small $m_a$. Note also that the harmonic time-dependencies of \eqref{eq:gRefraction} and \eqref{eq:gSquaredRefraction} depend only on the axion mass, not on the frequency of incident light, $p^0$.

\begin{figure}[h]
    \centering
    \includegraphics[scale=1]{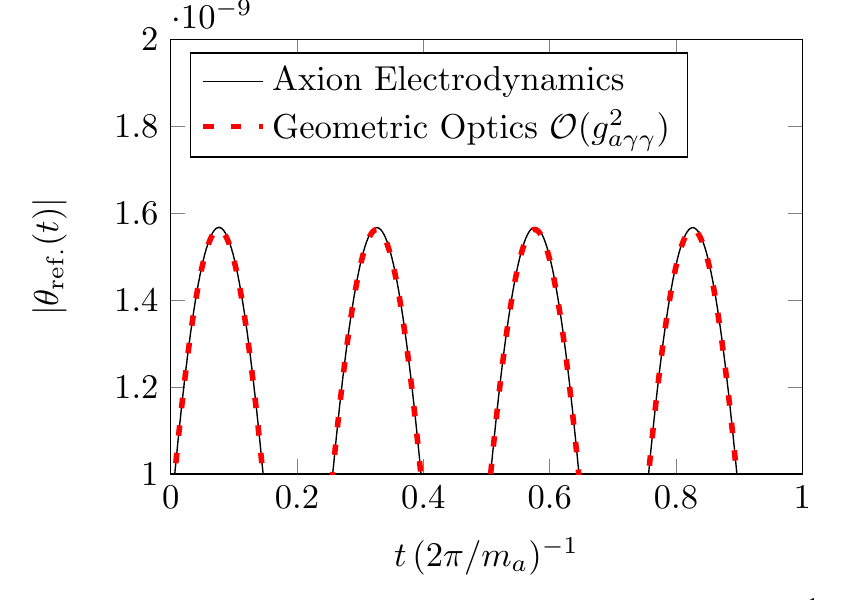}
    \caption{Evolution of the refraction angle (in rad) as a function of time in the region of parameter space where geometric optics is valid.  The black line was computed in full axion electrodynamics using the procedure in sec.~\ref{sec:numerics} whilst the red dashed line corresponds to eq.~\eqref{eq:theta}. The axion profile responsible for refraction is $a(t,x) = \sin(m_a t)e^{-x^2/L_a^2}$. The values chosen were were $L_a \omega_0 = 16$, $m_a/\omega_0 = 0.005$, $g_{a \gamma \gamma} a_0=0.04$ with $\omega_{\rm p}=0$. The incidence angle was taken to be $\theta_i = 0.02$. We chose a circularly polarised incident beam with $\boldsymbol{\varepsilon}_{0}=(-\sin \theta_i,\, i,\, \cos \theta_i) E_0$. }
    \label{fig:ThetaofT}
\end{figure}

\begin{figure}[h]
    \centering
    \includegraphics[scale=1]{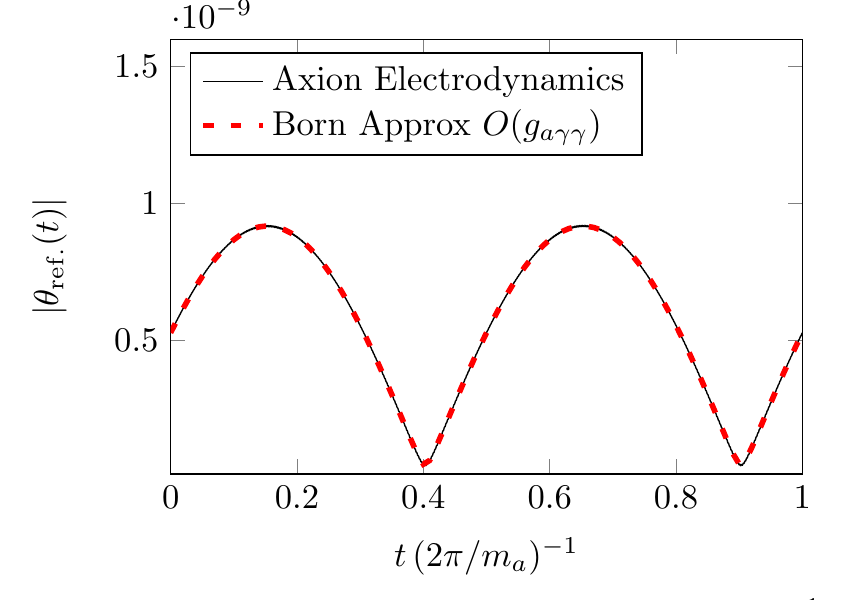}
    \caption{Plot of wave-optical refraction angle (in rad) - the black line corresponds to the numerical result, whilst the red dashed corresponds to the Born approximation of eq.~\eqref{eq:ABorn}. In this plot we took $g_{a \gamma \gamma} a_0 = 10^{-4}$ with $\omega_{\rm p}=0$. The incidence angle was taken to be $\theta_i = \pi/3$. The other values are the same as for fig.~\ref{fig:ThetaofT}.}
    \label{fig:ThetaofT2}
\end{figure}

\begin{figure*}[t!]
    \includegraphics[width = 0.95\textwidth]{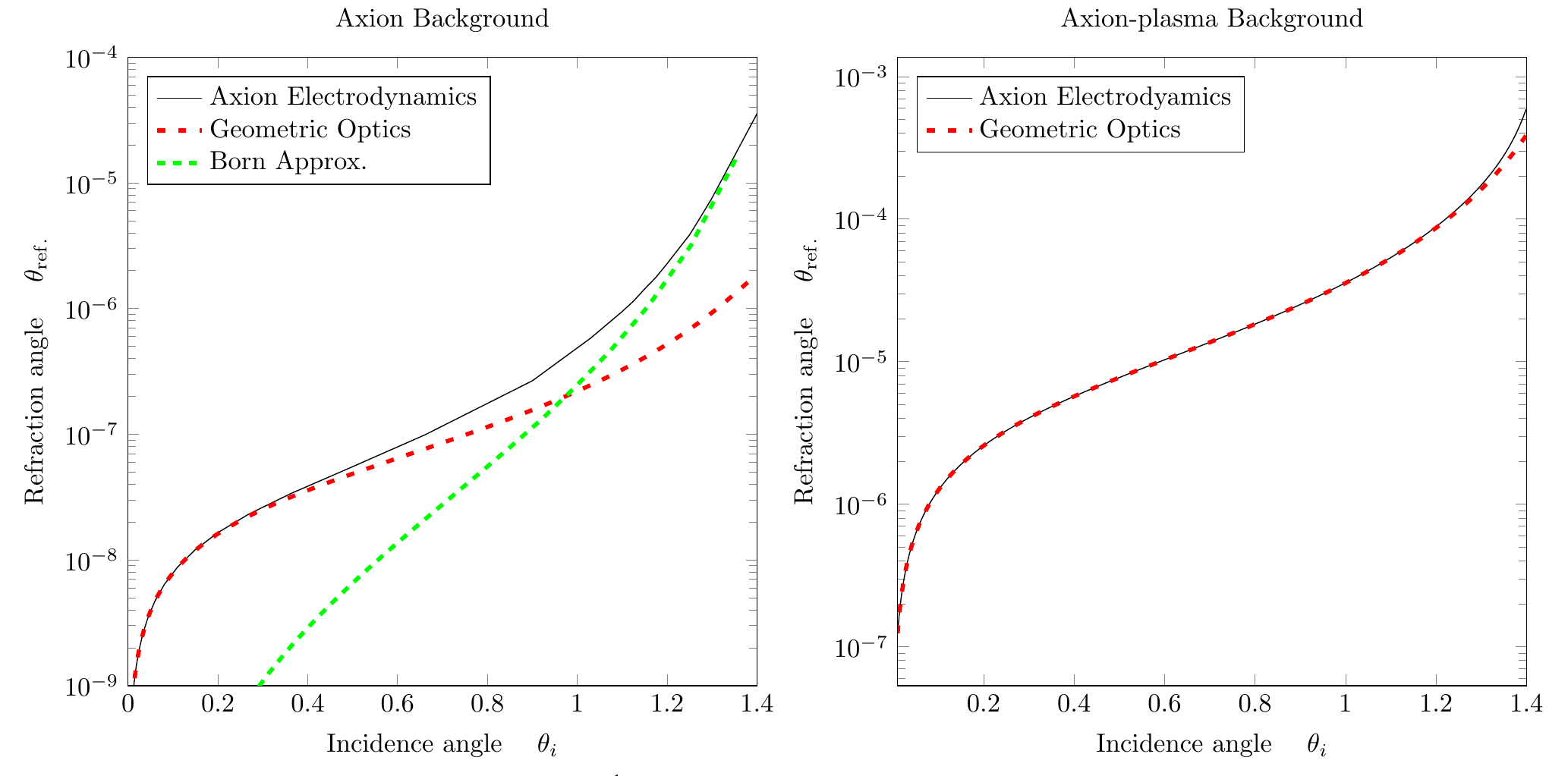}
    \caption{Refraction angle (in rad) as a function of incidence angle (in rad) without (left panel) and with plasma (right panel). The parameter values chosen illustrate the qualitative behaviour of the theory. For a phenomenological application consistent with current constraints see ref.~\cite{Prabhu:2020pzm}. In both cases, we took an axion profile $a(t,x) = \sin(m_a t)e^{-x^2/L_a^2}$, with $L_a \omega_0 = 16$, $m_a \omega_0 = 0.005$, $g_{a \gamma \gamma} a_0=0.04$, and chose a circularly polarised incident beam with $\boldsymbol{\varepsilon}_{0}=(-\sin \theta_i,\, i,\, \cos \theta_i)E_0$. For the plasma mass case we took $\omega_{ \rm p}  =0.6 \omega_{0}$. In each panel, the black curves indicate the solutions of full axion-electrodynamics computed up to the appropriate order in perturbation: left panel
    $\mathcal{O}(g_{a \gamma \gamma}^2)$ and right panel $\mathcal{O}(g_{a \gamma \gamma})$. The red-dashed curve shows the geometric optics formulae \eqref{eq:theta} (left panel) and \eqref{eq:thetaPL} (right panel). The green curve in the left panel shows the $\mathcal{O}(g_{a \gamma \gamma})$ Born approximation of eqs.~\eqref{eq:EandBBOrn}-\eqref{eq:ABorn} for comparison.}
    \label{fig:RefvsInce}
\end{figure*}

\section{Results}\label{sec:results}

We now come to the main point of this paper, the comparison of refraction in the full theory of axion electrodynamics derived in secs.~\ref{sec:Born} and \ref{sec:numerics} with the geometric optics approximation of sec.~\ref{sec:GeometricOptics}, in the limit where the photon wavelength is small compared to characteristic axion scales.

The refraction angle in the full theory is simply the angle between the incoming and outgoing Poynting fluxes
\begin{equation}
    \cos \theta_{\rm ref.}  = \hat{\textbf{S}}_{\rm in.} \cdot \hat{\textbf{S}}_{\rm out.},
\end{equation}
where $ \hat{\textbf{S}}_{\rm in.}$ and $\hat{\textbf{S}}_{\rm out.}$ are the unit vectors associated with the Poynting vectors of the incoming and outgoing beam. This can then be  compared with the geometric optics results \eqref{eq:theta}-\eqref{eq:thetaPL}.

The outgoing Poynting flux can be expanded as a perturbative expansion of terms $\mathcal{O}\left(g_{a \gamma \gamma}^n\right)$
\begin{equation}
\textbf{S}_{\rm out} = \sum_{n=0}^\infty  \textbf{S}^{(i)},
\end{equation}
where $\textbf{S}^{(0)} = \textbf{S}_{\rm in}$ and
\begin{align}
 \textbf{S}^{(1)}_{\rm out} &= \textbf{E}^{(0)} \times \textbf{B}^{(1)} +\textbf{E}^{(1)} \times \textbf{B}^{(0)}, \nonumber  \\
  \textbf{S}^{(2)}_{\rm out} &= \textbf{E}^{(0)} \times \textbf{B}^{(2)} +\textbf{E}^{(2)} \times \textbf{B}^{(0)} + \textbf{E}^{(1)}\times \textbf{B}^{(1)},
\end{align}
and so on, for higher orders, 
The results are plotted in figs.~\ref{fig:ThetaofT}-\ref{fig:RefvsInce}. In fig.~\ref{fig:RefvsInce}, in the absence of plasma, the electrodynamics solutions were computed numerically up to $\mathcal{O}(g_{a \gamma \gamma}^2)$ using the Born series described in sec.~\ref{sec:numerics}. For the plasma case, we plot the leading $\mathcal{O}(g_{a \gamma \gamma})$ contribution of axion-electrodynamics against the refraction angle predicted by geometric optics.

\subsection{Axion-only background}
We see that, in the absence of plasma, for small incidence angles and small deflection angles, the agreement between the $\mathcal{O}(g_{a \gamma \gamma}^2)$ refraction angle predicted by geometric optics and full-axion electrodynamics is good (see left part of left panel of fig. \ref{fig:RefvsInce}). For large incidence angles, geometric optics breaks down and the refraction angle is instead well-approximated by the analytical $\mathcal{O}(g_{a \gamma \gamma})$ Born approximation \eqref{eq:ABorn}. 



One might expect that latter phenomena could be suppressed relative to geometric optics by taking the photon momentum sufficiently large compared to the axion characteristic momentum scale, i.e. taking $|\mathbf{p}| \gg m_a$: that is, for large enough $|\mathbf{p}|$, one could always recover geometric optics. However, we can derive an explicit expression for the refraction angle predicted by the Born approximation in the limit $|\mathbf{p}|  \gg m_a$. This reads (appendix \ref{Appendix:BornAngle})
\begin{align}\label{eq:SmallmaBornAngle}
    &\sin \theta_{\rm ref.}  \simeq \frac{g_{a \gamma \gamma}}{2}\, \frac{m_a^3}{\left|\textbf{p}\right|^2} \,\frac{ \tan^3 \theta_i}{\cos \theta_i} \, \tilde{\psi}(0) \, \sin \left( m_a t \right). 
\end{align}
Comparing this expression with eq.~\eqref{eq:gSquaredRefraction} and fixing the ratio of axion spatial scales $L_a$ and frequencies $\sim m_a$ by taking $L_a m_a$ to be constant, the refraction angles associated with geometric optics and the Born approximation have the same frequency suppression, both scaling as $m_a^2/|\mathbf{p}|^2$. This is somewhat expected as the geometric optics result is associated with terms of the form $(\partial a)^2$ whilst the full theory necessarily incorporates terms $\partial_\mu \partial_\nu a$. Thus, once the short wavelength limit is taken, it is the direction of the ray relative to the optical axis which determines the regime of validity of geometric optics, rather than the value of frequency. 



\subsection{Axion-plasma backgrounds}

\begin{figure}
    \centering
    \includegraphics{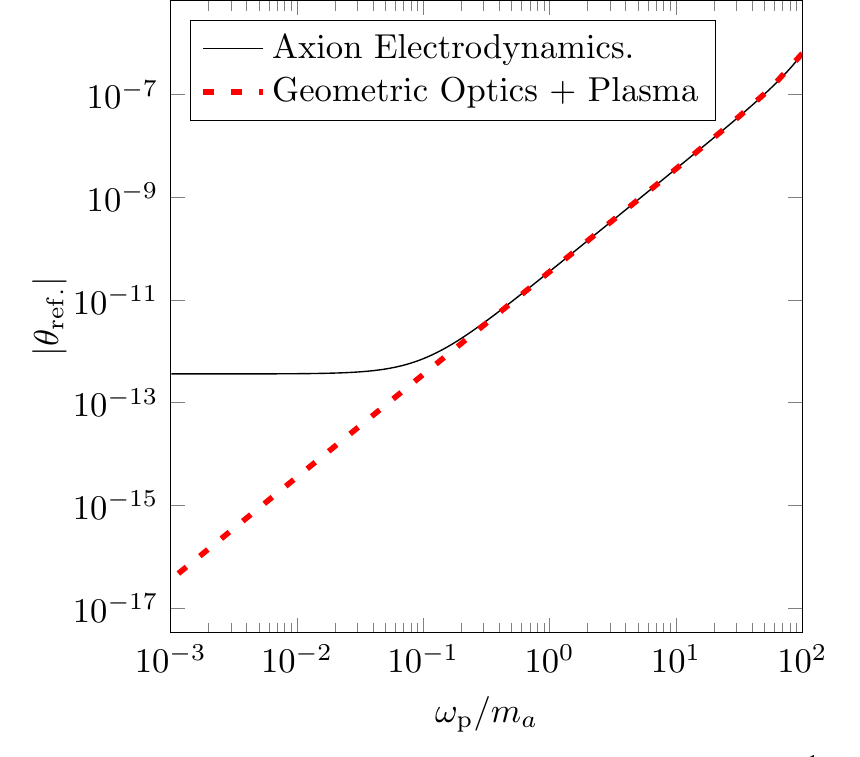}
    \caption{Dependence of refraction angle (in rad) on plasma mass - comparison of geometric optics (eq.~\eqref{eq:thetaPL}) and full axion electrodynamics. Other values are as in previous figures.}
    \label{fig:RefractionAngleVsPlasma}
\end{figure}

Turning now to the plasma case, we see that the agreement between geometric optics and the full theory is very good across a wide range of incidence angles, $\theta_i$.  A natural interpretation goes as follows. Since the photon acquires an effective mass, it has a Compton wavelength giving the photon a characteristic size.  When the curvature scales are larger than the characteristic size of the photon, as set by $1/\omega_{\rm p}$, the wave-nature of the photon becomes less important and tidal forces are suppressed. This is illustrated in fig.~\ref{fig:RefractionAngleVsPlasma}. However, for smaller plasma frequencies, the photon becomes highly relativistic and essentially massless, and the agreement between geometric optics and the full theory breaks down once more as curvature effects start to dominate. 

\section{Interpretation and other work}\label{sec:interpretation}

We now clarify some of the results presented here and their contrast with other discussions in the literature, as well as providing some interpretation for the above results. 

Starting with the former, we firstly stress that our treatment of geometric optics did not go beyond the eikonal approximation and we did not therefore include any kind of transport equation. Rather we derived the dispersion relation which, in general, can be expanded as a power series in $g_{a \gamma \gamma}$ \cite{Carroll:1989vb,Adshead:2020jqk,McDonald:2019wou,Prabhu:2020pzm}. The corrections in each order should be considered as a Born-like expansion in powers of $g_{a \gamma \gamma}$ about an unperturbed straight-line reference ray within the framework of the Hamiltonian optics, which admits perturbative solutions to the equations \eqref{dxdt}-\eqref{domegadt}. Thus, by building on the work of ref.~\cite{Blas:2019qqp}, higher order corrections to the dispersion relation are included as well as their corresponding corrections to the rays, without ever going beyond the leading order of WKB. 

Of course, at  some order in perturbation theory, the higher order $g_{a \gamma \gamma}^n$ corrections to the dispersion relation and refraction, must be smaller than those contributions which are not captured by geometric optics, e.g. higher order gradients and wave-optical effects of the full theory. Indeed this is precisely the point illustrated by fig.~\ref{fig:RefvsInce} which demonstrates the various regions of applicability. 

Secondly, we should also compare our treatment of geometric optics to that given recently in ref.~\cite{Schwarz:2020jjh}. 
Here, the authors essentially took the photon frequency to infinity as in \cite{Landau:1982dva}. This result therefore provided an alternative derivation of the polarisation rotation of light \cite{Harari:1992ea,Carroll:1989vb,Fujita:2018zaj,Alighieri:2010eu,Liu:2019brz,Fedderke:2019ajk,Basu:2020gsy}, which is frequency-independent and therefore remains non-negligible in the limit of zero wavelength. 

However, our interest is in the refraction of rays. Whilst the zero wavelength limit might be natural for gravitational lensing, ray refraction is typically frequency-dependent, so rays carry information about frequency of the photon. Instead, we retain small, but \textit{finite} wavelengths. Thus one can still meaningfully discuss rays, provided the small wavelength (i.e. WKB) limit applies, such that a local plane-wave ansatz is valid. The worldlines of these rays are then described by the set of Hamilton's equations. Furthermore, as we have shown, this ray picture matches with the direction of Poynting flux in the appropriate regime of validity. This is precisely what is done in e.g. refs.~\cite{Sluijter:s,WeinbergWKB}. 



This is a generic property arising from the fact that the refractive index of optical backgrounds typically depends on frequency. A classic example would be something like a refractive index $n = \sqrt{\varepsilon \mu_0}$ given by a Drude permittivity $\varepsilon(\omega) = 1 - \omega_{\rm p}^2/(\omega^2 - i \omega \Gamma)$, where $\Gamma$ is some dissipation factor. This can then be expanded as a power series in inverse frequency 
such that refraction is describable in terms of a power series in $1/\omega$. Taking the strict limit $\omega \rightarrow \infty$ would remove this physics completely. In fact, many optical properties are typically power series in $\omega'/\omega$ where $\omega'$ is some background frequency or momentum scale. 


This should be distinguished from the kind of refraction which happens at a sharp interface between two optical media (e.g. glass and air), which cannot be described by a WKB approximation and Hamiltonian optics. Here, the gradients vary over sub-wavelength scales and so the whole concept of a local momentum space and controlled gradient expansion breaks down, and Hamiltonian optics ceases to be meaningful. See the excellent discussion in \cite{Sluijter:s} which distinguishes these two scenarios. 


Secondly, we offer an interpretation of the breakdown of geometric optics with larger incidence angles, though caution is urged since a wider study of other axion geometries is clearly necessary in order to understand the generality of the following arguments. 

One possibility is that rays which have large incidence angles $\theta_i$, lie further from the optical axis and are more sensitive to the \textit{local} curvature of the background, $ \sim \partial^2 a$ (a breakdown in the slowly-varying-background assumption of Hamiltonian optics). Here the $\mathcal{O}(g_{a \gamma \gamma})$  become just as important as the $\mathcal{O}(g_{a \gamma \gamma}^2)$ corrections to Hamiltonian optics, $ \sim (\partial a)^2$.
We also note that giving the photon a finite mass $\omega_{\rm p}$ appears to improve the agreement with geometric optics. This would again hint at a suppression of the background curvature effects for sufficiently small Compton wavelengths. This can be seen as shrinking the effective size $\sim 1/\omega_{\rm p}$ of the photon to the point that it is no longer vulnerable to tidal forces.

Nonetheless, the most interesting and novel result of the present paper is that wave-optical refraction occurs at $\mathcal{O}(g_{a \gamma \gamma})$ and that this is not suppressed by additional powers of $\omega$ relative to the $\mathcal{O}(g_{a \gamma \gamma}^2)$ geometric optics result. A complete theory of lensing which includes wave optics effects should therefore be developed.

\section{Conclusions}\label{sec:conclusions}


In this paper, we vindicated our prior claim of ref.~\cite{McDonald:2019wou}, showing that when geometric optics is valid, light is refracted at $\mathcal{O}(g_{a \gamma \gamma}^2)$ and that this occurs for weakly lensed rays close to the optical axis. Of particular interest is the fact that light is still scattered at $\mathcal{O}(g_{a \gamma \gamma})$ and that this leads to \textit{wave-optical} refraction at $\mathcal{O}(g_{a \gamma \gamma})$ as measured by the Poynting flux of outgoing radiation. This analysis warrants a further discussion of refraction which goes beyond geometric optics.

Since we dealt with a very minimal axion background setup, the present analysis should be extended to include a wider variety of axion backgrounds, with e.g. spherical symmetry and a formalism for dealing with wave-optical lensing. The range of techniques developed in this paper (especially those in sec.~\ref{sec:numerics}) will prove useful in comparing any such theory against the predictions of full electrodynamics.

Once a complete theory of refraction has been derived, it would be interesting to re-examine the analysis of optical lensing by axion stars first presented in \cite{Prabhu:2020pzm} which used our previous geometric optics formula \eqref{eq:theta} derived in ref.~\cite{McDonald:2019wou}. In addition the analysis of polarisation-dependent light-bending by superradiant black holes \cite{Plascencia:2017kca} should be re-run.  The interesting lensing patterns due to the presence of scalar hair around black holes \cite{Cunha:2015yba, Cunha:2016bjh,Vincent:2016sjq} is another avenue of interest. In our case, this would arise from  \textit{direct} coupling of the scalar to photons, rather than by its gravitational potential.

\section*{Acknowledgements}
J.I.M acknowledges the support of the Alexander von Humboldt foundation and L.B.V is supported by FCT grant PD/BD/140917/2020 and by the CIDMA Project
No.~UID/MAT/04106/2020. We thank Joerg Jaeckel, Gonzalo Alonso-Álvarez, David J.E. Marsh, G{\"u}nter Sigl, Pranjal Trivedi, Andrea Caputo and Javier Redondo for useful conversations. We are also grateful to Dominik Schwarz, Jishnu Goswami, Aritra Basu and Bj{\"o}rn Garbrecht for correspondence and comments on draft manuscripts. Finally we thank Jan Sch{\"u}tte-Engel and Stefan Knirck for discussions on boundary conditions in electrodynamics. This research has also benefited from discussions held at the Munich Institute for Astro- and Particle Physics (MIAPP) which is funded by the Deutsche Forschungsgemeinschaft (DFG, German Research Foundation) under Germany's Excellence Strategy – EXC-2094 – 390783311.

\appendix

\section{Refraction in the Born Approximation}\label{Appendix:BornAngle}

In the case where the incident photon energy is greater than the axion mass $p^0 > m_a$, the possible outgoing energies are $(p^0 \pm m_a) > 0$, giving rise to two contributions in eq.~\eqref{eq:ABorn} consisting of momenta $p^\pm$ given by eq.~\eqref{eq:PLR}. For an incident circularly polarised wave, the initial polarisation vector is
\begin{equation}\label{eq:incCircular}
    \varepsilon_{\rm 0}  = E_0 \left(  - \sin \theta_i\,,  i,\, \cos \theta_i \right),
\end{equation}
where $E_0$ is the amplitude of the incident electric field. For simplicity, let us carry out the calculation in the absence of plasma. Then the outgoing momenta can be written as
\begin{equation}
    \textbf{p}^{\pm} = |\textbf{p}^\pm|(\cos  \theta^\pm , 0 , \sin \theta^\pm),
\end{equation}
where $\theta^\pm$ gives the angle between the outgoing momentum vectors $\textbf{p}^\pm$ and the incident momentum vector $\textbf{p}$. After a little algebra, the outgoing contributions to the gauge field then take the form
\begin{align}\label{eq:outcircular}
 \textbf{A}^\pm  =& \frac{i g_{a \gamma \gamma} E_0}{ \cos \theta^\pm} \sin^2 \left(\frac{\Delta \theta^\pm}{2}\right) \nonumber \\
 &
 \cdot 
  \left(
 \begin{array}{c}
     \sin \theta^\pm \\
       i  \\
     -\cos \theta^\pm
 \end{array}
 \right)  \tilde{\psi}\left (\Delta p^\pm \right) 
  e^{- i p^{\pm \, 0}  t + i \textbf{p}^\pm \cdot \textbf{x} },
\end{align}
where
\begin{equation}
\Delta \theta^\pm = \theta^\pm  - \theta_i, \qquad \Delta p^\pm = p'_x - p_x ,
\end{equation}
give the relative angle between the incoming and outgoing momenta and the momentum transfer from the axion field, respectively. Our goal is to compute the angle between the incoming wave and the outgoing Poynting flux. Note that one can treat $\tilde{\psi}(\Delta p^{\pm})$ as real, since any complex phase can be absorbed into the phase $p^{\pm \, 0}  t +  \textbf{p}^\pm \cdot \textbf{x}$ in the exponent and, as shown below, this results only in a phase shift of the final answer. 

Since the modes are both circularly polarised, with the outgoing fields polarised in the opposite sense to the incident one, there is a simple relation between electric and magnetic fields
\begin{equation}\label{eq:circularproperty}
\textbf{B}_{\rm inc } = i \textbf{E}_{\rm inc}, \qquad    \textbf{B}^\pm = - i \textbf{E}^\pm ,
\end{equation}
where $\textbf{E}$, $\textbf{E}^\pm$ and $\textbf{B}^\pm$ are the electromagnetic fields associated to eqs.~\eqref{eq:incCircular} and \eqref{eq:outcircular}. In other words, for circularly polarised fields, the electric and magnetic fields are related through a rotation of $\pi/2$.

Since we want to know the angle in which the outgoing radiation propagates, we are interested in computing the Poynting flux at $\mathcal{O}(g_{a \gamma \gamma})$. This is the sum of two contributions
\begin{equation}\label{eq:S1}
  \textbf{S}^{(1)}_{\pm} = \text{Re}\left[ \textbf{E}_{\rm inc} \right] \times  \text{Re}\left[ \textbf{B}_{\pm} \right] +  \text{Re}\left[ \textbf{E}_\pm \right] \times  \text{Re}\left[ \textbf{B}_{\rm inc} \right].
\end{equation}
We then expand the real parts of eq.~\eqref{eq:S1} in terms of the electromagnetic fields and their hermitian conjugate, making extensive use of the relations \eqref{eq:circularproperty}. After several cancellations, one is then left with a simple expression for the $\mathcal{O}(g_{a \gamma \gamma})$ contribution to the Poynting flux in terms of complex amplitudes
\begin{equation}\label{eq:S1EB}
     \textbf{S}^{(1)}_{\pm} = \frac{1}{2}\textbf{E}_{\rm inc} \times \textbf{B}^\pm + \text{h.c.} \, .
\end{equation}
Next we shall perform two rotations which make the determination of \eqref{eq:S1EB} considerably easier. Firstly, we rotate about the $y$-axis clockwise by an angle $\theta_i$
\begin{align}
&\textbf{B}^\pm \rightarrow \textbf{R}_y(\theta_i) \cdot \textbf{B}^\pm,
\qquad 
&\textbf{E}_{\rm inc}    \rightarrow \textbf{R}_y(\theta_i)\cdot \textbf{E}_{\rm inc} . 
\end{align}
The rotated fields then read
\begin{align} 
 \textbf{B}^\pm  =& \frac{i g_{a \gamma \gamma}|\textbf{p}^\pm| E_0}{ \cos \theta^\pm} \sin^2 \left(\frac{\Delta \theta^\pm}{2}\right) \nonumber \\
 &
 \cdot 
  \left(
 \begin{array}{c}
    \sin \Delta \theta ^\pm \\
       i  \\
     -\cos \Delta \theta^\pm
 \end{array}
 \right)  \tilde{\psi}\left (\Delta p^\pm \right) 
  e^{- i p^{\pm \, 0}  t + i \textbf{p}^\pm \cdot \textbf{x} } \label{eq:Brot},\\
  \nonumber \\
 \textbf{E}_{\rm inc}  &= \left(0, i , 1 \right) e^{- i p^0  t + i \textbf{p} \cdot \textbf{x} } . \label{eq:Erot}
\end{align}
In this frame we then have an incdient Poynting flux
\begin{equation}
\textbf{S}^{(0)} = \left(E^2_0,0,0 \right),
\end{equation}
and after a little algebra, combining eqs.~\eqref{eq:Brot} and \eqref{eq:Erot} gives the first order Poynting flux \eqref{eq:S1EB}
\begin{align}
   \textbf{S}^{(1)}_{\pm} = &  \frac{i |\textbf{p}^\pm|g_{a \gamma \gamma} E^2_0\sin^2 \left(\Delta \theta^\pm/2\right)}{\cos \theta^\pm  } \nonumber \\ 
  & \cdot   \left(
 \begin{array}{c}
     (1 + \cos \Delta \theta^\pm) \cos \chi^\pm \\
       -\sin \Delta \theta^\pm \sin \chi^\pm  \\
     \sin \Delta \theta^\pm \cos \chi^\pm
 \end{array}
 \right)  \tilde{\psi}\left (\Delta p^\pm \right) ,
\end{align}
where
\begin{equation}
\chi^\pm =  \left(p^{\pm \, 0} t +  \textbf{p}^\pm \cdot \textbf{x}\right) + \left(p^{0} t +  \textbf{p} \cdot \textbf{x}\right),
\end{equation}
results from combining the phases of the incident and scattered fields in eq.~\eqref{eq:S1EB}. Thus, in this frame, the incident Poynting flux simply points in the $x$-direction, and the two scattered contributions $\textbf{S}^{(1)}_\pm$ to the outgoing radiation have precession-like motion. Finally we can compute the angle between the outgoing Poynting flux
\begin{equation}\label{eq:Sout}
\textbf{S}_{\rm out} = \textbf{S}^{(0)} + \textbf{S}^{(1)}_{+} + \textbf{S}^{(1)}_{ -},
\end{equation}
and incoming Poynting flux
\begin{equation}\label{eq:Sin}
    \textbf{S}_{\rm in} = \textbf{S}^{(0)} =(E^2_0,0,0) .
\end{equation}
To do so in a simple fashion, we perform our second transformation, which consists of a time-dependent rotation by one of the phases $\chi^\pm$ about the $x$-axis. This of course leaves all the angles in the problem unchanged. Let us perform a rotation by $-\chi^+$ about the $x$-axis
\begin{align}
\textbf{S}^{(0)} \rightarrow \textbf{R}_x( - \chi^+) \cdot \textbf{S}^{(0)} , \qquad    \textbf{S}^{(1)}_{\pm} \rightarrow \textbf{R}_x( - \chi^+) \cdot   \textbf{S}^{(1)}_{\pm} .
\end{align}
Clearly the incident Poynting flux is invariant under a rotation about the $x$-axis since it is aligned with it. After performing this transformation we therefore have
\begin{widetext}
\begin{align}
  \textbf{S}^{(1)}_+ &=   \frac{i |\textbf{p}^+|g_{a \gamma \gamma} E^2_0\sin^2 \left(\Delta \theta^+/2\right)}{ \cos \theta^+  }    \left(
 \begin{array}{c}
     (1 + \cos \Delta \theta^+) \cos \chi^+ \\
      0 \\
     \sin \Delta \theta^+
 \end{array}
 \right)  \tilde{\psi}\left (\Delta p \right) \label{eq:Splus} , \\
  \textbf{S}^{(1)}_- & =   \frac{i |\textbf{p}^-|g_{a \gamma \gamma} E^2_0\sin^2 \left(\Delta \theta^-/2\right)}{\cos \theta^-  }    \left(
 \begin{array}{c}
     (1 + \cos \Delta \theta^-) \cos \chi^- \\
      -\sin(\chi^- - \chi^+) \sin \Delta \theta^-\\
     \cos(\chi^- - \chi^+) \sin \Delta \theta^-
 \end{array}
 \right)  \tilde{\psi}\left (\Delta p^- \right)  \label{eq:Sminus}, \\
 \textbf{S}^{(0)} &= (E_0^2,0,0). \label{eq:S0}
\end{align}
\end{widetext}
We are now in a position to compute the refraction angle in a compact way. This is nothing more than the angle between the vectors in eq.~\eqref{eq:Sout} and \eqref{eq:Sin}. From the eqs.~\eqref{eq:Splus}-\eqref{eq:S0} above, it is easy to see that this angle is given by simple trigonometry:
\begin{equation}
\sin \theta_{\rm ref.} = \frac{\sqrt{ (S_{\rm out})^2_y + (S_{\rm out})^2_z}}{\sqrt{ (S_{\rm out})^2_x + (S_{\rm out})_y^2 + (S_{\rm out})^2_z}},
\end{equation}
To leading order in $g_{a \gamma \gamma}$ this just gives
\begin{equation}\label{eq:sintheatref}
\sin \theta_{\rm ref.}   \simeq \frac{ \sqrt{ \left[S^{(1)}_{+ \, y}+S^{(1)}_{- \, y} \right]^2 +\left[S^{(1)}_{+ \, z}+S^{(1)}_{- \, z} \right]^2  } }{E_0^2},
\end{equation}
from which it easy to see that to leading order in $g_{a \gamma \gamma}$, the refraction angle only depends on time through the combination,
\begin{equation}
\chi^+ - \chi^- = 2 m_a t + (\textbf{p}^+ - \textbf{p}^-)\cdot \textbf{x}. 
\end{equation}
That is, the refraction angle is independent of incident photon frequency, $p^0$, oscillating with a frequency set solely by the axion mass. Hence without loss of generality, we can set $\textbf{x} =0$, which is just equivalent to an overall phase shift.

We can then finally expand the following quantities up to leading order in $m_a$
\begin{align}
&\sin \Delta \theta^\pm \simeq \mp \frac{m_a}{p^0} \tan \theta_i, \nonumber \\
&\theta^\pm \simeq \theta_i , \nonumber \\
&\tilde{\psi}(\Delta p^\pm )\simeq \tilde{\psi}(0), \nonumber \\
&|\textbf{p}'^\pm| \simeq p^0.
\end{align}
Inserting these small $m_a$ approximations into \eqref{eq:Splus}, \eqref{eq:Sminus} and \eqref{eq:sintheatref}, leads, after keeping only the leading order contributions in powers of $m_a$ to the expression
\begin{equation}
    \sin \theta_{\rm ref.}  \simeq \frac{g_{a \gamma \gamma}}{2}\, \frac{m_a^3}{\left|\textbf{p}\right|^2} \,\frac{ \tan^3 \theta_i}{\cos \theta_i} \, \tilde{\psi}(0) \, \sin \left( m_a t \right),
\end{equation}
quoted in eq.~\eqref{eq:SmallmaBornAngle} in the main text.


\bibliography{References}

\end{document}